\documentclass[10pt,twocolumn,aps,prb,floatfix,superscriptaddress]{revtex4-2}
\usepackage{graphicx}
\usepackage{color}
\usepackage{amsmath}
\usepackage{amssymb}
\usepackage{comment}
\usepackage{hyperref}

\begin{document}
\title{The positively charged carbon vacancy defect as a near-infrared emitter in 4H-SiC}
\date{\today}

 \author{Meysam Mohseni}
 \affiliation{Wigner Research Centre for Physics, PO.\ Box 49, H-1525 Budapest, Hungary}
 \affiliation{Lor\'and E\"otv\"os University, P\'azm\'any P\'eter S\'et\'any 1/A, H-1117 Budapest, Hungary}
 \author{P\'eter Udvarhelyi}
\affiliation{Wigner Research Centre for Physics, PO.\ Box 49, H-1525 Budapest, Hungary} 
\affiliation{Budapest University of Technology and Economics, M\H{u}egyetem rkp.\ 3., 1111 Budapest, Hungary}

 \author{Gerg\H{o} Thiering}
\affiliation{Wigner Research Centre for Physics, PO.\ Box 49, H-1525 Budapest, Hungary} 
 \author{Adam Gali}
\affiliation{Wigner Research Centre for Physics, PO.\ Box 49, H-1525 Budapest, Hungary} 
\affiliation{Budapest University of Technology and Economics, M\H{u}egyetem rkp.\ 3., 1111 Budapest, Hungary}

\begin{abstract}
Certain intrinsic point defects in silicon carbide are promising quantum systems with efficient spin-photon interface. Despite carbon vacancy in silicon carbide is an elementary and relatively abundant intrinsic defect, no optical signal has been reported associated with it. Here, we revisit the positively charged carbon vacancy defects in the 4H polytype of silicon carbide (4H-SiC) by means of \textit{ab initio} calculations. We find that the excited state is optically active for the so-called h-site configuration of carbon vacancy in 4H-SiC, with zero-phonon line at $0.65~\mathrm{eV}$. We propose this defect as an exotic paramagnetic near-infrared emitter in the IR-B region.
\end{abstract}

\maketitle

\section{Introduction}

Silicon carbide (SiC) is one of the most promising material platforms for the integration of quantum defects. Wafer scale single crystals and isotope engineered samples are readily available. Scalable arrays of integrated quantum emitters were already demonstrated in this platform~\cite{Radulaski2017, Wang_2017}. Advanced micro-fabrication techniques and recent advances in the integrated photonic devices~\cite{Bracher_2017, Song_2019, Lukin2020, Guidry_2020} enable for improved magneto-optical properties for the hosted quantum emitters. Various point defects in the 4H polytype of SiC have been already utilized for quantum communication~\cite{Lohrmann2015}, quantum computing~\cite{Weber_2011} and nanoscale sensing~\cite{Kraus2014, Falk_2014, Simin2015, Cochrane2016} applications. Most of the promising defects are vacancy-related, created by irradiation techniques~\cite{Hemmingsson_1997, Chakravorty_2021} or laser writing~\cite{Liu2021}.
Isolated silicon vacancy ($\text{V}_\text{Si}$) is a well known fundamental defect in silicon carbide associated with single photon emitters with coherently controllable electron spin even at room temperature~\cite{Sorman_2000, Janzen_2009, Nagy_2018, Nagy_2019, Viktor2018, Kraus_2014, Widmann2015, Niethammer_2019}. Divacancy ($\text{V}_\text{Si}\text{V}_\text{C}$) is a defect complex of adjacent silicon and carbon vacancies~\cite{Carlos_2016} created by annealing. Its neutral charge state is a color center~\cite{Falk_2013, Gali_2010, Ivadi2017} with a triplet ground state for which coherent manipulation was firstly demonstrated among the quantum defects in SiC~\cite{Koehl_2011, Christle_2015}. For the other constituent of the divacancy, the single carbon vacancy defect, no associated emission was reported in experiments, which is a critical information for the initialization and readout of the spin state. As the electron irradiation creation of this fundamental defect is even more efficient than that of the silicon vacancy~\cite{Steeds_2001}, we investigate its possible application as an abundant quantum defect in SiC.

The carbon vacancy ($\text{V}_\text{C}$) defect in 4H-SiC was investigated by various experimental techniques before. Deep level transient spectroscopy (DLTS)~\cite{Danno2005} and electron paramagnetic resonance (EPR)~\cite{Son2001} measurements assigned two paramagnetic centers, EI5 and EI6, to the $\text{V}_\text{C}^{+}$ charge state corresponding to the quasi-cubic (k) and hexagonal (h) defect sites, respectively.
The four Si neighbours of the defect showed considerable hyperfine constants. The EI5 center was reported to be Jahn-Teller (JT) distorted to $\text{C}_{1\text{h}}$ symmetry~\cite{Umeda2004e}. Temperature activated averaging to $\text{C}_{3\text{v}}$ symmetry was also reported above T $\approx50$~K, with an activation energy of $0.014~\mathrm{eV}$~\cite{Coutinho2017,Umeda2004e}. From photo-EPR measurements, Son et al.~\cite{Son2002} revealed that the EI5 center is a stable deep donor, with $(+/0)$ charge transition level at 1.47~eV above the valence band maximum~\cite{Danno2005}. However, no optical signal has been associated with these basic defects in 4H-SiC which is surprising as photoluminescence is a very sensitive technique when compared to DLTS or EPR methods.

The $\text{V}_\text{C}$ defect was investigated in several theoretical studies too. The formation energy was calculated for both k- and h-site to be $4.07~\mathrm{eV}$ and $4.21~\mathrm{eV}$, respectively~\cite{Torpo2001,Hornos2011}. Hyperfine constants calculated with LSDA method showed good agreement with experimental data~\cite{Ayedh2015, Coutinho2017}. The Jahn-Teller distortion in the k-site defect was revealed to be pseudo-Jahn-Teller (pJT) effect, originating from the electron-phonon interaction between $a$ and $e$ defect orbitals in the non-degenerate ground state~\cite{Coutinho2017}. The lack of pJT effect for the h-site was attributed to the larger crystal field around the core of the defect.

In this work, we determine the key optical and spin properties of the positively charged carbon vacancy ($\text{V}_\text{C}^{+}$) defect at both k- and h-sites in 4H-SiC by means of advanced density functional theory calculations. We confirm its identification to the EI5 and EI6 EPR centers. Our calculations reveal a zero-phonon line (ZPL) at the h-site defect in the near-infrared wavelength region associated with a relatively large Debye-Waller factor. This provides crucial information for the optical initialization of the quantum state, a first step towards the coherent control of its spin. Based on these findings, we propose the h-site $\text{V}_\text{C}^{+}$ defect as a promising quantum emitter possessing a paramagnetic ground state.


\begin{figure*}
\includegraphics[scale=.42]{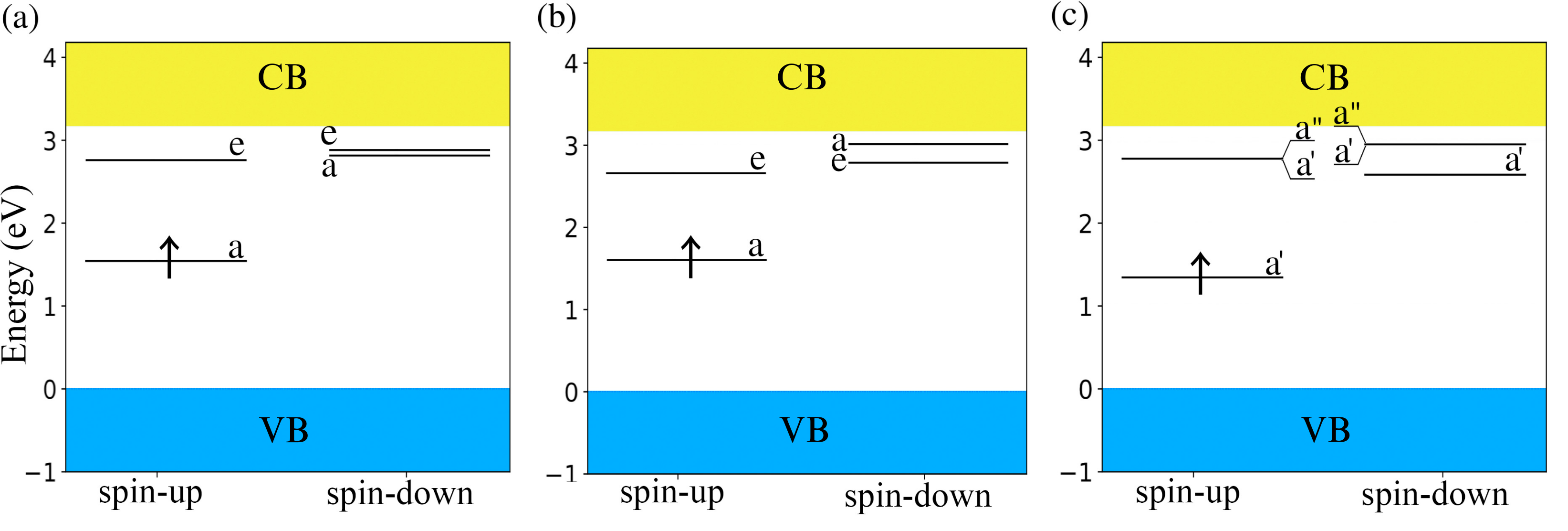}
\caption{
Electronic level diagrams of the $\text{V}_{\text{C}}^{+}$ defect in $\text{C}_{3\text{v}}$ symmetry at (a) h-site and (b) k-site and (c) in $\text{C}_{1\text{h}}$ symmetry at k-site. Each Kohn-Sham level is labeled according to the irreducible representations of the corresponding point group symmetry. The conduction band (CB) and the valance band (VB) are shown in green and blue colors, respectively.
}
\label{fig:levels}
\end{figure*}

\section{Methods}
The electronic properties of $\text{V}_{\text{C}}^{+}$ were calculated using density functional theory (DFT)~\cite{PhysRev.140.A1133,PhysRev.136.B864} as implemented in the Vienna Ab-initio Simulation Package (VASP) plane wave based  code~\cite{PhysRevB.54.11169,Kresse1996}. The hybrid exchange functional of Heyd, Scuseria,and Ernzerhof (HSE06) was used~\cite{doi:10.1063/1.1564060}.
The defect structure is modeled in a $6\times 6\times 2$ supercell (576 atoms), allowing for accurate $\Gamma$-point sampling in the k-space.
The atomic positions were optimized until the Hellman-Feynman forces acting on them were less than 0.01~eV/\AA .
$\Delta$SCF or constraint occupation DFT~\cite{Gali2009} was employed for the calculation of the electronic excitations. For the calculation of phonon spectrum and normal modes, the functional of Perdew, Burke and Ernzerhof (PBE)~\cite{PhysRevLett.77.3865} was applied using density functional perturbation theory (DFPT) method. Choosing the exchange correlation functional of PBE over HSE06 in our phonon calculations provides reasonably accurate results while significantly reducing the computational overhead. Transition dipole matrix elements were calculated from the overlap of the pseudo wavefunctions between the Kohn-Sham (KS) states involved in the excitation, as implemented in the PyVaspwfc code~\cite{Zheng}. Nonradiative transition rates were calculated with the NONRAD code of Turiansky {\it et al.}~\cite{alkauskas_first-principles_2014, turiansky_nonrad_2021}.

Partially self-consistent GW0~\cite{Shishkin_2007} and BSE~\cite{BSE1, BSE2} calculations were performed using VASP in a 128-atom supercell model, keeping the $\Gamma$-point sampling. The orbital basis was calculated with DFT HSE06 functional, with number of unoccupied orbitals more than 15-times the occupied ones. The energy cutoff for the calculation of the response function was limited to $100~\mathrm{eV}$. BSE was calculated beyond the Tamm-Dancoff approximation~\cite{Sander_2015}, including 50 occupied and 50 unoccupied orbitals.

\section{Results}
\begin{figure}
\centering
\includegraphics[scale=0.07]{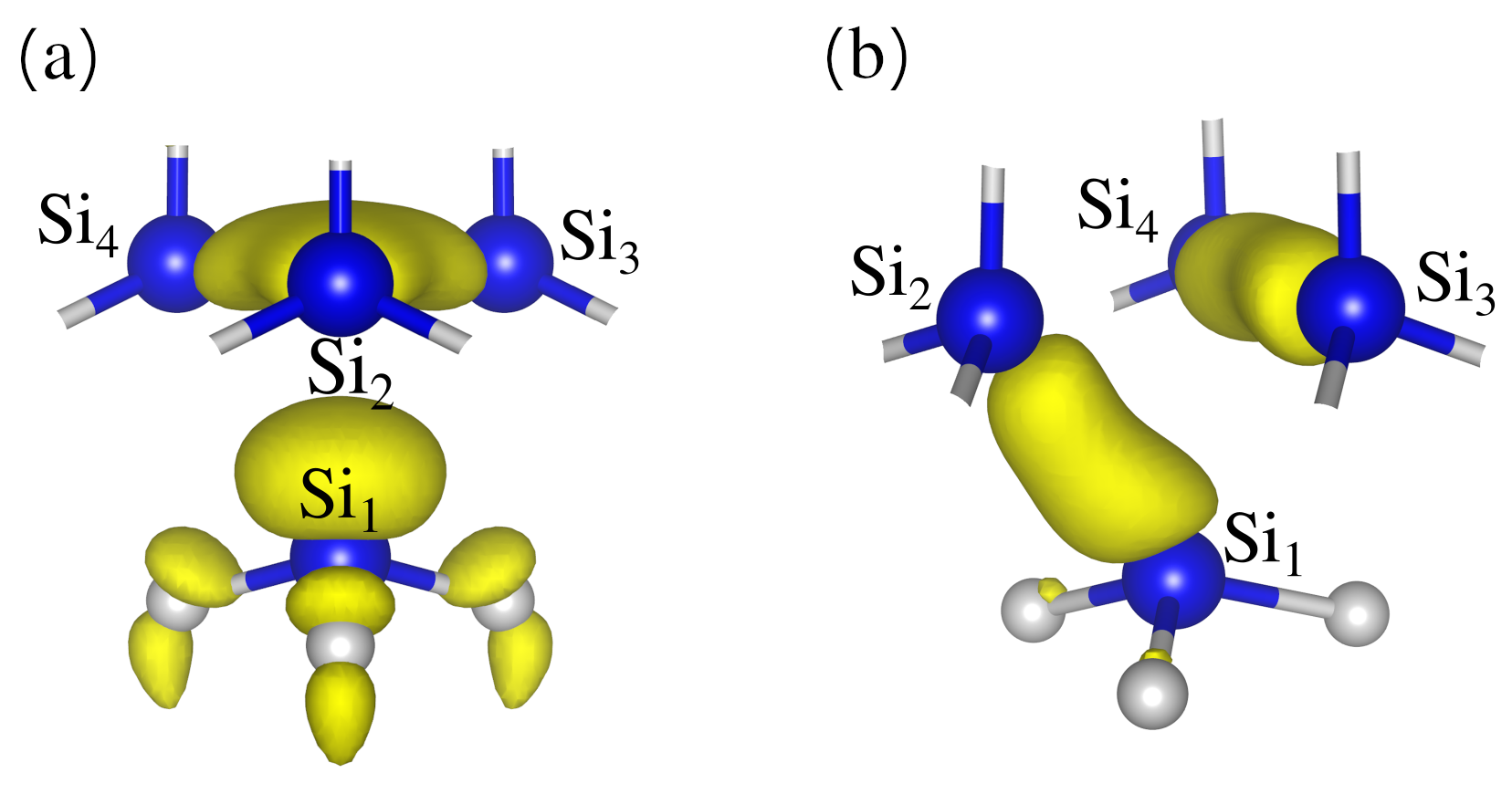}
\caption{
Spin density of the V$_{\text{C}}^{+}$ defect at (a) h-site with $\text{C}_{3\text{v}}$ symmetry and (b) k-site with $\text{C}_{1\text{h}}$ symmetry. The hyperfine parameters for the atoms labeled here are given in Table~\ref{T3}.
}
\label{fig:hf}
\end{figure}

The carbon vacancy model in 4H-SiC exhibits $\text{C}_{3\text{v}}$ symmetry without considering the electron-phonon interaction (by restricting the symmetry). The vacancy dangling bonds introduce two defect levels, $a$ and $e$, in the band gap, labeled by the irreducible representations of the symmetry group. In the positively charged ground state of the defect, a single electron occupies the $a$ level, while the double degenerate $e$ level is empty, resulting in a spin doublet state. After lifting the symmetry constraints, we obtain a pJT distorted relaxed structure at k-site with $\text{C}_{1\text{h}}$ symmetry with a pJT relaxation energy of $82~\mathrm{meV}$. However, the same type of calculations at the h-site results in a negligible pJT energy, in line with previous findings~\cite{Coutinho2017}.  
The defect-level-diagram for both site defects is shown in Fig.~\ref{fig:levels}. In Fig.~\ref{fig:hf}, we visualize the geometric structure and the ground state spin density for both h- (a) and k-sites (b). For the $\text{C}_{3\text{v}}$ symmetric h-site defect, the distance between the three top-side silicon atoms ($\text{Si}_2$, $\text{Si}_3$ and $\text{Si}_4$) and their distance to $\text{Si}_1$ are $3.08~\text{\AA}$ and $3.22~\text{\AA}$, respectively. For $\text{C}_{1\text{h}}$ symmetry at k-site, the distance between Si$_1$-Si$_2$ and Si$_3$-Si$_4$ are $3.15~\text{\AA}$ and $3.02~\text{\AA}$, respectively. The Si$_{3,4}$-Si$_2$ and Si$_{3,4}$-Si$_1$ distances are $3.10~\text{\AA}$ and $3.26~\text{\AA}$, respectively.

\subsection{Hyperfine parameters}

\begin{table*}
\caption{\label{T3} 
HSE06 calculated and experimental~\cite{Umeda2004} hyperfine principal values ($A_{xx}$, $A_{yy}$ and $A_{zz}$) for the nearest neighbor silicon atoms (see Fig.~\ref{fig:hf}) in the $\text{V}_{\text{C}}^{+}$ defect. The core contributions is included in the results. $\theta$ is the polar angle of the $A_{zz}$ principal axis, measured from the c-axis of the 4H-SiC crystal.
}
\begin{ruledtabular}
\begin{tabular}{ccccccccccccc}
 &&&HSE06 calculation&&&&&&Experimental&&&\\
  & Atoms & $A_{xx}~(\mathrm{MHz})$ & $A_{yy}~(\mathrm{MHz})$ & $A_{zz}~(\mathrm{MHz})$&$\theta~(^\circ)$&&&$A_{xx}~(\mathrm{MHz})$ & $A_{yy}~(\mathrm{MHz})$ & $A_{zz}~(\mathrm{MHz})$&$\theta~(^\circ)$\\ \hline
h-site &Si$_1$& -323.3 & -323.3 &  -492.4 &0&&& 297.29 & 297.29 & 433.75&0  \\
& Si$_2$-Si$_3$-Si$_4$ &  -17.1 &  -15.0 & -30.5  &90.33&&& 39.23 & 39.23 &  59.12&98 \\
\\
k-site  & Si$_1$  & -130.7	 & -127.2 &  	-198.4&8.7&&& 124.4 & 94.14 & 181.01&7.7  \\
& Si$_2$ & -67.1 &  -64.7 & -105.5&125.79&&&  91.06 & 89.38 &  132.81&121.5  \\
&Si$_3$-Si$_4$ & -99.45 & -96.3 &-148.6 &102.23&&& 107.88 & 106.76 & 154.67&103.2
\label{hft}
\end{tabular}
\end{ruledtabular}
\end{table*}
For the EPR-active doublet ground state of the V$_{\text{C}}^{+}$ defect, we calculate the hyperfine parameters using the HSE06 hybrid functional. The hyperfine tensor describing the interaction between the nuclear spin at $R_I$ and the electron spin density $\rho_s$ of the defect is given by
\begin{equation}
A_{ij}=\frac{4\pi}{3}\frac{g_N\gamma_Ng\gamma_e}{\left<\hat S_z\right>}\int \mathrm{d}^3r \rho_s(r)m_{i,j}\left(r-R_I\right)\text{,}
\end{equation}
where $g_N$, $\gamma_N$, $g$ and $\gamma_e$, are the $g$ factors and gyromagnetic ratios of the nucleus and the electron, respectively. The $m_{i,j}=[\delta_{ij}\delta(r)-\frac{1}{2}({3x_ix_j-r^2\delta_{i,j}}){r^{-5}}]$ is the interaction potential
between electron and nuclear spins, which include the Fermi-contact and a dipole-dipole interaction terms.
Our results for the hyperfine parameters of the first neighbor silicon atoms are listed in Table~\ref{hft}, at both k- and h-sites. We also compare the results to experimental data for EI5 (k-site and C$_{\text 1h}$ symmetry) and EI6 (h-site and C$_{\text 3v}$ symmetry)~\cite{Umeda2004}, showing reasonable agreement.

\subsection{Vibronic spectrum}

Next, we describe the vibronic interaction in the k-site ground state and model the microscopic origin of thermal averaging to $\text{C}_{3\text{v}}$ symmetry, observed in the EI5 EPR center~\cite{Umeda2004e}. 
It was attributed to the dynamic pJT effect in the ground state in Ref.~\onlinecite{Coutinho2017}; however, their calculated barrier energy was considerably larger ($50~\mathrm{meV}$) than the thermal activation energy of $14~\mathrm{meV}$. 
In our HSE06 calculations, we obtain $76~\mathrm{meV}$ for the barrier energy, which is comparable to the pJT energy itself ($82~\mathrm{meV}$). Based on these results, we cannot consider the pJT barrier as a perturbation directly corresponding to the dynamics of the system. Instead, we go beyond the Born-Oppenheimer approximation and approach the problem as a dephasing process through strongly coupled phonon excitations of the JT active modes. In this model, we identify the thermal occupation of the first vibronic excited state of the system as the onset of the dynamics, where the activation energy corresponds to the polaronic gap, i.e., the energy difference of the first vibronic excited state and the vibronic ground state.

\begin{figure*}
\includegraphics*[scale=.65]{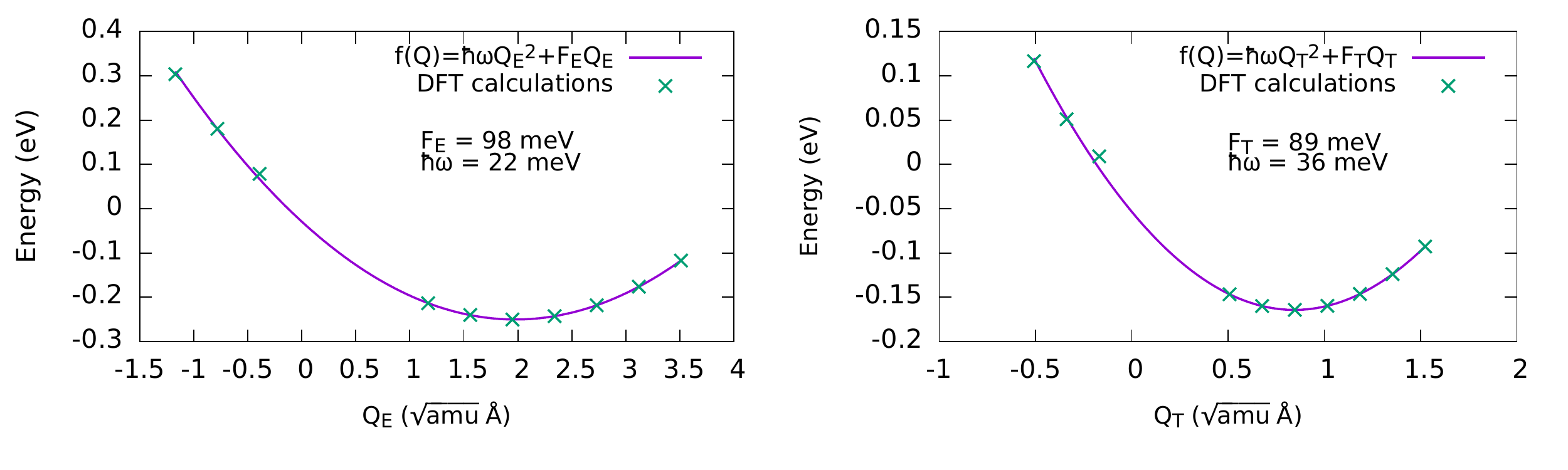}
\caption{Adiabatic potential energy surfaces in the function of configuration coordinates corresponding to (a) $T$- and (b) $E$-symmetric distortions in the V$_{\text{C}}^{+}$ defect at k-site. The fitted curve corresponds to the linear Jahn-Teller solution of the $T\times(e+t_\text{2})$ problem.
}
\label{fig:PES}
\end{figure*}

To this end, we apply a similar model as detailed in Ref.~\onlinecite{Udvarhelyi2020}. We separate the effect of the crystal field present in the 4H-SiC crystal and describe the defect orbitals in high symmetry $\text{T}_\text{d}$. In this picture, the vacancy dangling bonds introduce a single $t_2$ orbital into the band gap, occupied by a single electron. This $^{2}T$ electronic configuration is Jahn-Teller unstable coupling to phonon modes of $t_2$ and $e$ symmetries, called the $T\otimes(e\oplus t_2$) problem. The $\text{C}_{3\text{v}}$ crystal field is added as a perturbation in this model.
The potential-energy surfaces (PES) of the $T$ orbitals are
formulated with pseudo-spin of three dimensions. Therefore, the vibronic interaction can be expressed on this basis as a $3\times 3$ matrix:
\begin{equation}
W = \begin{bmatrix}
F_E\left(\frac{Q_{\vartheta}}{2}-\frac{\sqrt{3}Q_{\varepsilon}}{2}\right) & -F_T Q_\zeta & -F_T Q_\eta\\
-F_T Q_\zeta & F_E\left(\frac{Q_{\vartheta}}{2}+\frac{\sqrt{3}Q_{\varepsilon}}{2}\right) & -F_T Q_\xi\\
-F_T Q_\eta & -F_T Q_\xi & -F_E Q_\vartheta
\end{bmatrix},
\end{equation}
where the orbital degrees of freedom ($t_2^{(x)}$, $t_2^{(y)}$ and $t_2^{(z)}$) are depicted by the rows and columns of the $3\times 3$ matrix and the vibrational degrees of freedom are expressed by the $Q_i$ configuration coordinates. $F_{\text{T}}$ and $F_{\text{E}}$ are the linear vibronic coupling parameters of the corresponding phonon symmetries.
The linear vibronic coupling can be expressed with the Jahn-Teller energy ($E_\text{JT}$) as [see Eq. (3.46, 3.48) in Ref. \onlinecite{Bersuker2006} ]
\begin{align}
F_E = \sqrt{2\hbar\omega_{E} E_\text{JT}^{E}},& &
F_T = \sqrt{\frac{3}{2}\hbar\omega_{T} E_\text{JT}^{T}}\text{,}
\end{align}
where $\hbar\omega$ is the phonon energy of the corresponding JT active mode in the harmonic approximation. Our calculations results for the determination of the above parameters can be seen in Fig.~\ref{fig:PES}, where we fit a linear vibronic coupling model for each JT modes. 
Thus, the adiabatic potential-energy surface (APES)
\begin{align}
\nonumber\mathbf{\varepsilon}(\mathbf{Q}) =& \frac{1}{2}\hbar\omega_\text{E} \left(Q_\varepsilon^2+Q_\vartheta^2\right){\textbf I}
+\frac{1}{2}\hbar\omega_\text{T}\left(Q_\xi^2+Q_\zeta^2+Q_\eta^2\right){\textbf I}\\
&
+\mathbf{W}(\mathbf{Q})-\frac{\delta}{3}
\begin{pmatrix}
0 & 1 & 1\\
1 & 0 & 1\\
1 & 1 & 0
\end{pmatrix}.
\end{align}
%
consists of the phonon energy associated to the harmonic
potential of the electronic APES (\textbf{I} is the identity matrix), the vibronic interaction, and the crystal-field splitting ($\delta$). The latter can be obtained by artificially turning off the origin of the occupational instability, i.e., by smearing the electron occupation on all the three $t_2$ defect orbitals. The residual splitting corresponds to the crystal field parameter $\delta=85~\mathrm{meV}$.
The associated Hamiltonian describes three coupled five-dimensional harmonic oscillators as
\begin{equation}
  \begin{aligned}
\hat{H} =&
\hbar\omega_{E} 
\left(\hat{a}_\varepsilon^\dagger \hat{a}_\varepsilon+
\hat{a}_\vartheta^\dagger \hat{a}_\vartheta+1\right)\\
&
+\hbar\omega_{T}\left(
\hat{a}_\xi^\dagger \hat{a}_\xi+
\hat{a}_\eta^\dagger \hat{a}_\eta +
\hat{a}_\zeta^\dagger \hat{a}_\zeta+
 \frac{3}{2}\right)\\
&
-F_E\left(\hat{T}_\varepsilon \hat{Q}_\varepsilon
+\hat{T}_\vartheta \hat{Q}_\vartheta\right)\\
&
-F_T\left(\hat{T}_\xi \hat{Q}_\xi
+\hat{T}_\eta \hat{Q}_\eta
+\hat{T}_\zeta \hat{Q}_\zeta\right)\\
&
-\frac{\delta}{3}\left(\hat{T}_\xi
+\hat{T}_\eta
+\hat{T}_\zeta\right)\text{,}
\end{aligned}
\end{equation}
\\
where $\hat{a}_i^\dagger$ is the oscillator $i$-mode creation operator, $\hat{Q}_i=\frac{1}{\sqrt{2}}(\hat{a}^\dagger_i+\hat{a}_i)$
are the coordinate operators, and the pseudo-spin of $T$ orbitals is represented by the orbital operators
\begin{equation}
  \begin{aligned}
&\hat{I}=\begin{pmatrix}
1 & 0 & 0\\
0 & 1 & 0\\
0 & 0 & 1
\end{pmatrix},
&
\hat{T}_\varepsilon=\begin{pmatrix}
\frac{\sqrt{3}}{2} & 0 & 0\\
0 & -\frac{\sqrt{3}}{2} & 0\\
0 & 0 & 0
\end{pmatrix},\\
&
\hat{T}_\vartheta=\begin{pmatrix}
-\frac{1}{2} & 0 & 0\\
0 & -\frac{1}{2} & 0\\
0 & 0 & 1
\end{pmatrix},
&
\hat{T}_\xi=\begin{pmatrix}
0 & 0 & 0\\
0 & 0 & 1\\
0 & 1 & 0
\end{pmatrix},\\
&
\hat{T}_\eta=\begin{pmatrix}
0 & 0 & 1\\
0 & 0 & 0\\
1 & 0 & 0
\end{pmatrix},
& 
\hat{T}_\zeta=\begin{pmatrix}
0 & 1 & 0\\
1 & 0 & 0\\
0 & 0 & 0
\end{pmatrix}.
\end{aligned}
\end{equation}

We solve the model Hamiltonian using the following ansatz as the basis of the vibronic states
\begin{equation}
\begin{aligned}
\left|\widetilde{\Psi}\right>=\sum_{j,k,l,n,m}&\left(c_{jklnm}^{(\varepsilon)}\left|t_{2}^{(\varepsilon)}\right> + c_{jklnm}^{(\vartheta)}\left|t_{2}^{(\vartheta)}\right>\right.\\&+c_{jklnm}^{(\xi)}\left|t_{2}^{(\xi)}\right> + c_{jklnm}^{(\eta)}\left|t_{2}^{(\eta)}\right> \\&\left. +c_{jklnm}^{(\zeta)}\left|t_{2}^{(\zeta)}\right>\right) \left|j,k,l,n,m\right>
\text{,}
\end{aligned}
\end{equation}
where $\mathcal{O}=\left(j+k+l+n+m\right)$ is the order of phonon excitations, acting as the cutoff for the basis size. We solve for the first polaronic excited state energy in the function of the excitation order up-to $\mathcal{O}=10$ and extrapolate the finite order basis assuming exponential convergence in the energy. The final result for the polaronic gap is $14.4~\mathrm{meV}$, in excellent agreement with the observed activation energy of the thermal averaging in the EI5 center.

\subsection{Excited state calculations}

In the following, we describe the electronic structure of the excited states. The lowest lying excited state of the defect can be described as promoting a single electron from the $a$ orbital to the $e$ degenerate orbital in the $\text{C}_{3\text{v}}$ symmetric ground state configurations. In this vertical excitation, the electron-hole interaction decreases the $a-e$ KS level gap. We obtain vertical excitation energies of $0.794~\mathrm{eV}$ and $0.967~\mathrm{eV}$ for the k- and h-site defects, respectively. The former is in good agreement with the GW+BSE result of $0.89~\mathrm{eV}$ reported in Ref.~\cite{Bockstedte2010}. Next, we allow the relaxation of the atomic positions in the restricted high symmetry. During the relaxation, the KS level gap is further decreased. This results in ZPL energies of $0.656~\mathrm{eV}$ and $0.839~\mathrm{eV}$ for the k- and h-site defects, respectively. However, these excited states are JT unstable, further relaxing to $\text{C}_{1\text{h}}$ symmetry. For the k-site defect, the effect of this relaxation changes the qualitative picture of the electron promotion, as the order of the occupied $e$ and empty $a$ level is interchanged. The resulting electronic structure indicates that the k-site defect does not form a stable emitter. This effect is not present at the h-site despite of the calculated JT energy of $0.187~\mathrm{eV}$. We account the stability of the h-site emitter to the larger crystal field splitting. Its final ZPL energy connecting the JT distorted excited state to the high-symmetry ground state is $0.652~\mathrm{eV}$.

We apply GW+BSE calculations in order to prove the stability of the optical emission from the JT distorted excited state of the h-site defect. In our partially self-consistent EVGW0 calculations in the ground state $\text{C}_{3\text{v}}$ geometry, we obtain quasi-particle levels of 1.93 eV, 2.76 eV, and 3.53 eV energies with respect to the VBM for the a, e, CBM levels, respectively. We note that the CBM level is obtained in the $\Gamma$-point of the 128-atom supercell which does not fold the $M$-point (CBM k-point). The vertical absorption in the BSE calculation is 1.029~eV. We apply the same methods in the JT distorted excited state geometry, resulting in the vertical emission at 0.187~eV. We obtain a 72\% contribution of the $a\rightarrow e$ transition in the excitonic wavefunction which implies that the constructed exciton wavefunction in the $\Delta$SCF procedure is relatively well described. The HSE06 relaxation energy between the optimised JT distorted excited state geometry and the ground state geometry is 0.339~eV in the ground state electronic configuration. The sum of this HSE06 relaxation energy and the BSE vertical emission energy results in an estimate for the ZPL energy of 0.526~eV, which is in good agreement with the ZPL energy obtained from the $\Delta$SCF method in a larger supercell. This result confirms the stability of the radiative transition in the h-site defect. 

\subsection{Photoluminescence spectrum}

\begin{figure}
\centering
\includegraphics[scale=0.65]{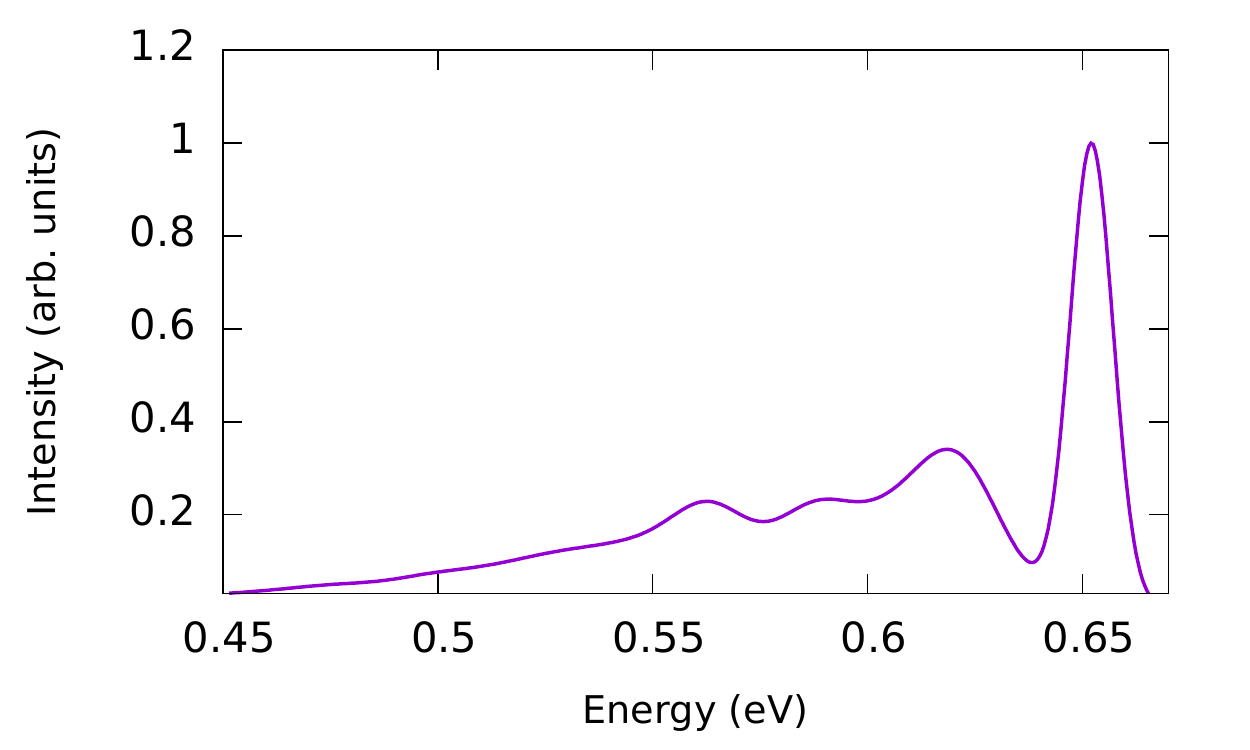}
\caption{ 
Simulated PL spectrum of the h-site $\text{V}_{\text{C}}^{+}$ defect. The calculated HR and DW factors are $1.605$ and $20\%$, respectively. }
\label{fig:PL}
\end{figure}

The photoluminescence line-shape is calculated using the method of Alkauskas {\it et al.}~\cite{Alkauskas_2014}.
The calculated PL spectrum from h-site V$_{\text{C}}^{+}$ defect is plotted in Fig.~\ref{fig:PL}.
The Huang-Rhys (HR) factor determines the intensity of the phonon sideband. It  quantifies the electronic and vibronic states. In one-dimensional approximation, the HR factor is given by
\begin{equation}
S=\frac{E_{\rm FC}}{\hbar\omega_{0}},
\end{equation}
where $\omega_{0}$ is the vibrational frequency of the effective mode and $E_{\rm FC}$ is the Frank-Condon relaxation energy~\cite{PhysRevB.102.134103}. The Debye-Weller (DW) factor is the ratio of the ZPL intensity in the total emission. It is directly related to the HR factor by
\begin{equation}
W_\text{ZPL}= e^{-S}\text{,}
\end{equation}
where S is the total HR factor. For our calculated $S=1.605$,  the DW factor is $20\%$.

\subsection{Radiative lifetime}
The transition dipole moment ($\mu$) between the
ground and excited state is calculated using the pseudo-wavefunctions of the defect Kohn-Sham levels $a$ and $e$ in the JT distorted excited state.
The radiative transition rate of the h-site defect is calculated as 
\begin{equation}
\frac{1}{\tau_r}=\frac{n\omega^3|\mu|^2}{3\pi \epsilon_0 \hbar c^3}\text{,}
\end{equation}
where $c$ is the speed of light, $\hbar \omega=0.652~\mathrm{eV}$ is the calculated ZPL energy, $\mu=29.8~\mathrm{D}$ is the optical-transition dipole moment, $n=2.647$ is the refractive index of 4H-SiC, and $\epsilon_0$ is the vacuum permittivity.
The resulting lifetime is $\tau_r=9.3~\mathrm{ns}$, which is comparable to the values reported for the negatively charged nitrogen-vacancy center ($\text{NV}^{-}$) in diamond~\cite{nv}. We also note that the JT distortion has a large effect on the optical transition dipole moment. It slightly distorts the wavefunction along with the geometry relaxation. However, the largest effect is the greatly decreased energy separation of the Kohn-Sham levels resulting in a large transition dipole moment. Performing the same calculation in the restricted $\text{C}_{3\text{v}}$ symmetric h- and k-site excited states, we obtain optical lifetimes of $198~\mathrm{ns}$ and $321~\mathrm{ns}$, respectively. The latter results are in line with previous GW plus Bethe-Salpeter equation calculations reporting negligible absorption in the $a$ to $e$ defect level transition~\cite{Bockstedte2010}.

\subsection{Nonradiative transition}
\begin{figure}
\includegraphics[scale=0.65]{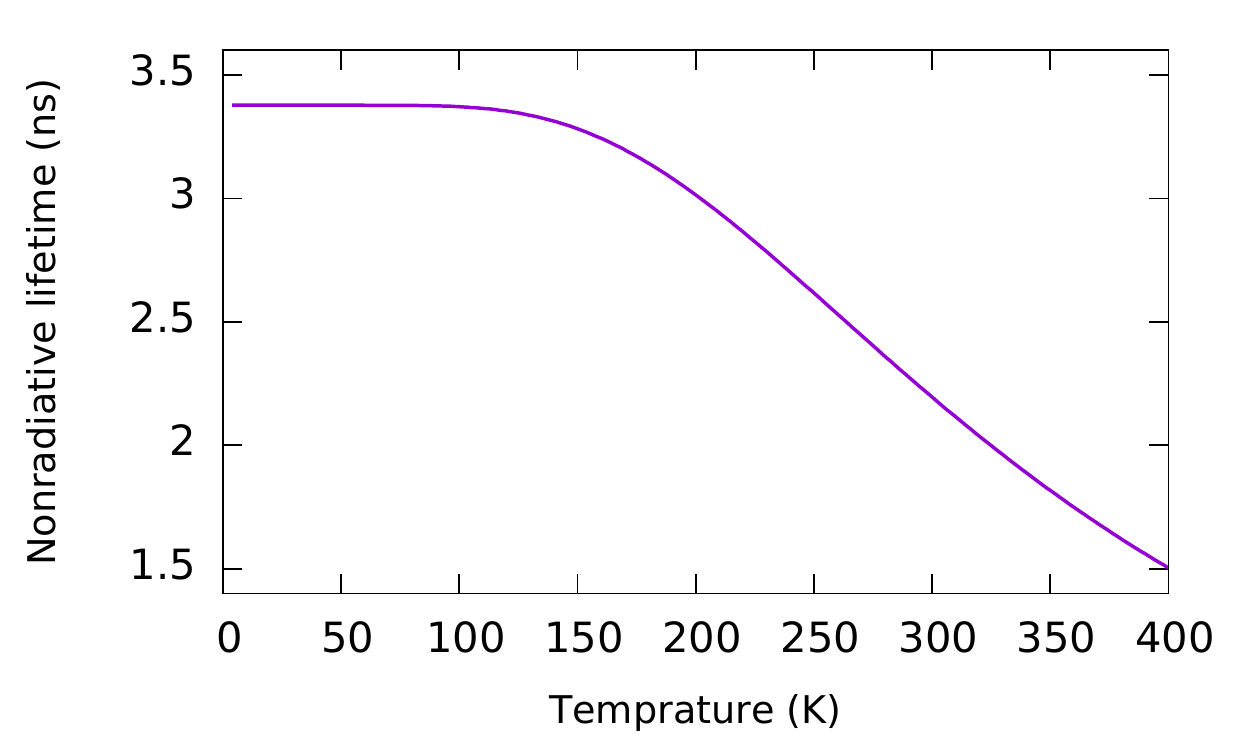}
\caption{Calculated nonradiative lifetime of the h-site $\text{V}_{\text{C}}^{+}$ defect in the function of temperature. 
 }
\label{fig:nonrad}
\end{figure}
Finally, we describe the nonradiative relaxation rate from the JT distorted excited state to the high symmetry ground state, coupled by electron-phonon interaction for the h-site carbon vacancy defect.
In this work, we consider the interaction within the first
order of electron-phonon coupling. According this assumption the capture rate is given by Fermi's golden rule~\cite{Stoneham2007};
\begin{equation}
r=\frac{2\pi}{\hbar}g\sum_{m}\omega_m \sum_{n} \left|\Delta H_{im;fn}^{e-ph}\right|^2 \delta(E_{im}-E_{fn})\text{,}
\end{equation}
where  $\omega_m$ is the thermal occupation of the vibrational state m of the excited electronic state, $E_{im}$ and $E_{fn}$ are total
energies of the initial and the final vibronic states, $g$ is the degeneracy factor of the final state. $\Delta H_{im;fn}^{e-ph}$ is the electron-phonon coupling matrix element which can be expressed as
\begin{equation}
\Delta H_{im;fn}^{e-ph}=\sum_{k} \left<\Psi_i\left|\partial H/\partial Q_k\right|\Psi_f \right>\left< \chi_{im}\left|Q_k-Q_{0;k}\right|\chi_{fn}\right>\text{,}
\end{equation}
where $H$ is Hamiltonian of the combined system of electrons and ions. The sum runs over all phonon modes $Q_k$ and $Q_{0;k}$ is the projection of the initial atomic configuration $Q_0$ along each of the phonon coordinates. $W_{if}=\left<\Psi_i\left|\partial H/\partial Q_k\right|\Psi_f \right>$ is the electron-phonon coupling matrix element pertaining to the phonon mode $k$.
In this work, the corresponding parameters are $g=3$, $W_{if}=0.15~\mathrm{eV}\mathrm{amu}^{-\frac{1}{2}} \text{\AA}^{-1}$, $\omega_a=0.05~\mathrm{eV}$ and $\omega_e=0.072~\mathrm{eV}$. The calculated nonradiative lifetime for the one-particle transition from $e$ to $a$ Kohn-Sham level is $\tau_{nr} \sim 3.38~\mathrm{ns}$. For this transition, the quantum efficiency (QE) of 27$\%$ was calculated as
\begin{equation}
\text{QE} = \frac{\tau_{nr}}{\tau_r + \tau_{nr}}\text{,}
\end{equation}
where $\tau_{nr}$ is nonradiative lifetime and $\tau_r$ is radiative lifetime. The calculated nonradiative lifetime at elevated temperatures is shown in Fig.~\ref{fig:nonrad}.

\section{Discussion}
The negatively charged carbon vacancy defect in 4H-SiC is a promising paramagnetic defect for quantum technology applications. Owing to its non-zero electron spin, its ground state structure have been already thoroughly investigated by {\it ab initio} calculations. Here we report the in-depth characterization of the excited state optical properties of the defect that have not been detected in experiments to-date. We find that the Jahn-Teller effect plays an important role in the stability of the excited state, enabling a stable $\text{C}_{1\text{h}}$ configuration at the h-site. We propose the $\text{V}_{\text{C}}^{+}$ defect as a near infrared (IR-B) emitter with the calculated ZPL energy of $0.652~\mathrm{eV}$. We find the properties of this emission very promising in the IR-B region. It shows a very short optical lifetime of $9.3~\mathrm{ns}$, which is excellent compared to the values of $11.6~\mathrm{ns}$, $14~\mathrm{ns}$ and $5.5~\mathrm{ns}$ values measured for $\text{NV}^{-}$ center in diamond~\cite{nv}, divacancy~\cite{Christle_2015} and negatively charged silicon vacancy~\cite{v1} in 4H-SiC, respectively. Moreover, the quantum efficiency of this transition is obtained to be 27\%. The optical coherence is remarkable as well, with the calculated Debye-Waller factor of $20\%$. We conclude that the carbon vacancy defect at the h-site in 4H-SiC can be observed in photoluminescence and may act as a single photon emitter when the defects are engineered to be isolated in the host material.

\section*{Acknowledgement}
 AG acknowledges the Hungarian NKFIH grant No.\ KKP129866 of the National Excellence Program of Quantum-coherent materials project and the support for the Quantum Information National Laboratory from the Ministry of Innovation and Technology of Hungary, and the EU H2020 project QuanTELCO (Grant No.\ 862721). The calculations were performed on the Hungarian Supercomputer Centre at KIF\"U.

\bibliography{SiC}
\end{document}